\documentclass[11pt]{article}
\usepackage{moriond}

\def\MET{\ensuremath{E_\mathrm{T}^{miss}}\ }
\def\HT{\ensuremath{H_\mathrm{T}}\ }
\def\TeV{TeV\ }
\def\GeV{GeV}
\def\pt{\ensuremath{p_\mathrm{T}}\ }
\def\char{\ensuremath{\widetilde{\chi}_1^+}\ }
\def\neut{\ensuremath{\widetilde{\chi}_2^0}\ }
\def\LSP{\ensuremath{\widetilde{\chi}_1^0}\ }

\newcommand{\Photo}{\includegraphics[height=35mm]{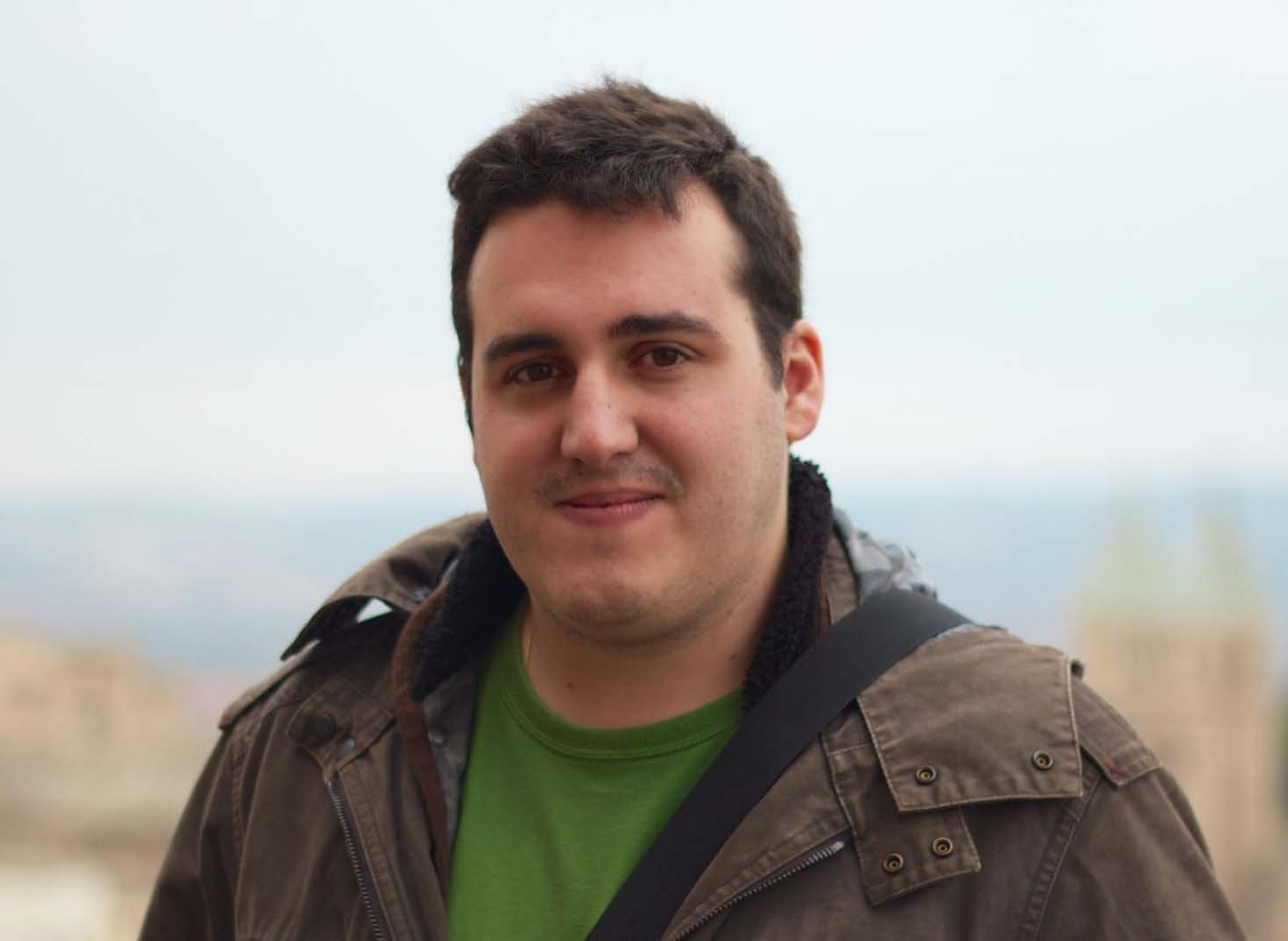}}

\begin{document}
\vspace*{4cm}
\title{SEARCH FOR NEW PHYSICS USING EVENTS WITH TWO SAME-SIGN ISOLATED LEPTONS IN THE FINAL STATE AT CMS}

\author{ S. Folgueras for the CMS Collaboration}

\address{Departamento de F\'isica, Universidad de Oviedo, \\
C/Calvo Sotelo s/n, Oviedo, Spain}

\maketitle\abstracts{We present a search for new physics using events with two same-sign isolated leptons with/out the presence of b-jets in the final state, targetting two very different SUSY scenarios, one dominated by strong production of squarks and gluinos where the $3^{rd}$ generation squarks are lighter than other squarks and the other dominated by electroweak production neglecting completely any strongly interacting particles. No excees above the standard model background is observed. The results are interpreted in various SUSY models.}

\section{Introduction}
Events with same sign dilepton final states are very rare in the SM context, but they appear naturally in many different new physics scenarios such as SUSY where two same-sign dileptons can be produced in the decay chain of supersymetric particles. 

Two different scenarios are considered: SUSY processes dominated by strong production of gluinos and squarks where $3^{rd}$ generation squarks are lighter than other squarks, resulting in an abundance of top and bottom quarks produced in the decay chain. Direct electroweak production of charginos(\char) and neutralinos(\neut), asuming that the strongly interacting particles are too heavy to play a role, resulting in events with multiple leptons in the final state. In either case the SUSY decay chain ends with the LSP (\LSP), that escapes undetected and therefore contribute strongly to the \MET of the event. 

In general, same-sign dileptons can be particularly sensitive to SUSY models with compressed spectra where the mass of the LSP is very close to the mass of the produced supersymetric particle, either if it is produced via strong production (squarks or gluinos) when it is accompanied with high hadronic activity or if it is produced via ewk production (charginos or neutralinos) when almost no hadronic activity is present.  We therefore search for SUSY using same sign dilepton events with/out hadronic activity and large \MET and we interpret the results in the context of various SUSY models. What we present here is just a short summary of two analysis performed  CMS \cite{CMS}, more details can be found in the original publications \cite{RA5,SUS12022}.

\section{Event selection}
We require two isolated same-sign leptons ($e$ or $\mu$) with \pt $>$ 20 \GeV, consistent with originating from the same vertex. Events are collected using dilepton triggers and an extra veto on the third lepton is applied to suppress Drell-Yan production. The isolation of the leptons is computed with particle-flow information, and an event-by-event correction is made to account for the effect of the multiple pp interaction in the same bunch crossing (pileup). This correction consists in substracting the estimated contribution from the pileup in the isolation cone. 

The baseline selection differs slightly depending which signature are we considering: strong (SS+b) or electroweak (EWK) production of SUSY. For the first we expect high bjet multiplicity so we will also require the presence of at least two b-tagged jets (with \pt $>$ 40 \GeV). For the second, hardly any hadronic activity is expected therefore we pick events with \MET $>$ 120\GeV (coming from the two LSP) to suppress background events. 
The signal regions are defined by impossing tighter cuts on the number of (b) jets, scalar sum of the \pt of all identified jets (\HT) and \MET for the analysis targeting strong production of SUSY, and on \MET and the number of bjets for the one targeting electroweak production.

\begin{figure} 
	\begin{center}
		\begin{minipage}{0.40\linewidth}
			\centerline{\includegraphics[width=0.9\linewidth]{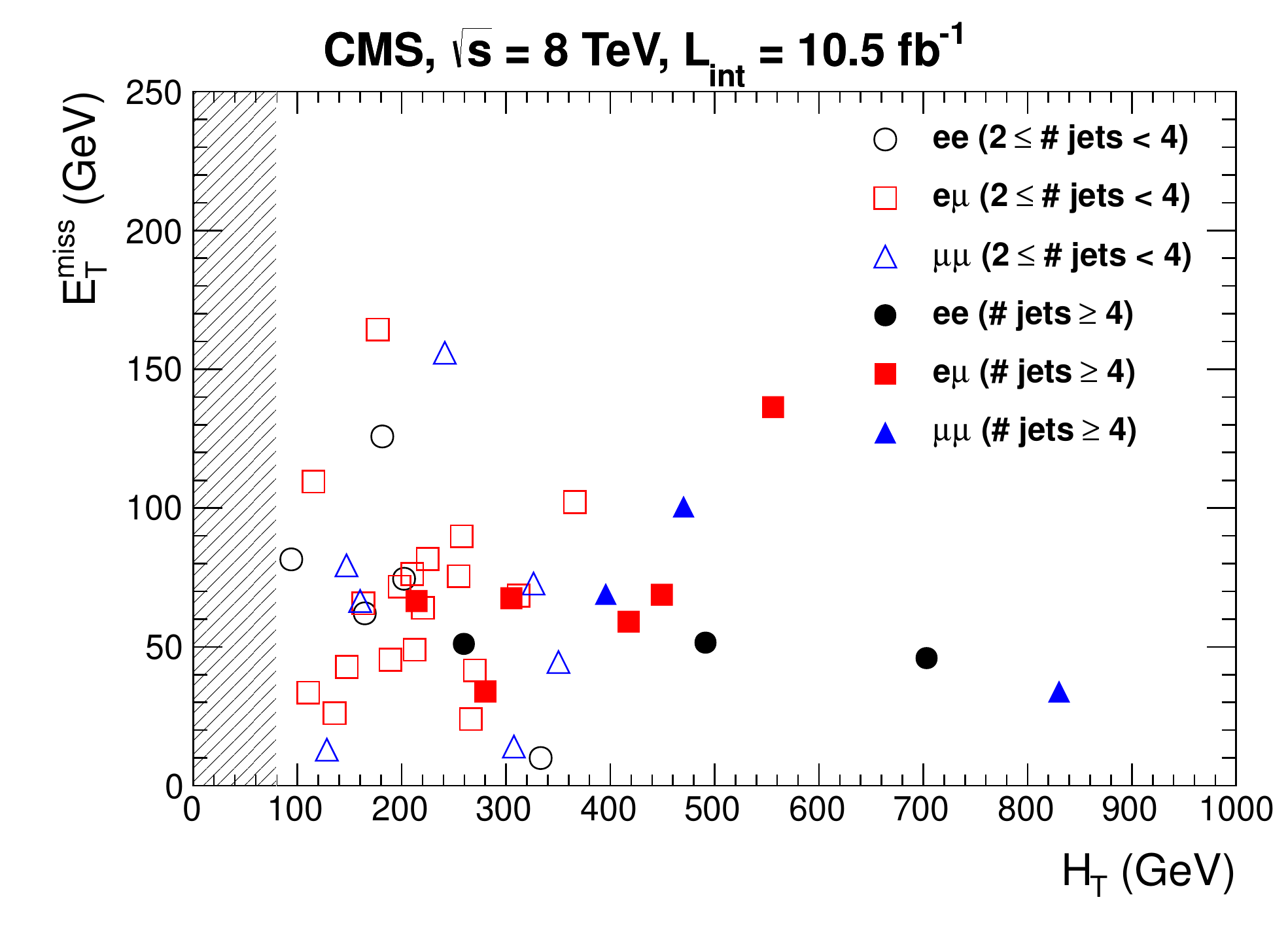}}
		\end{minipage}
		\begin{minipage}{0.28\linewidth}
			\centerline{\includegraphics[width=0.9\linewidth]{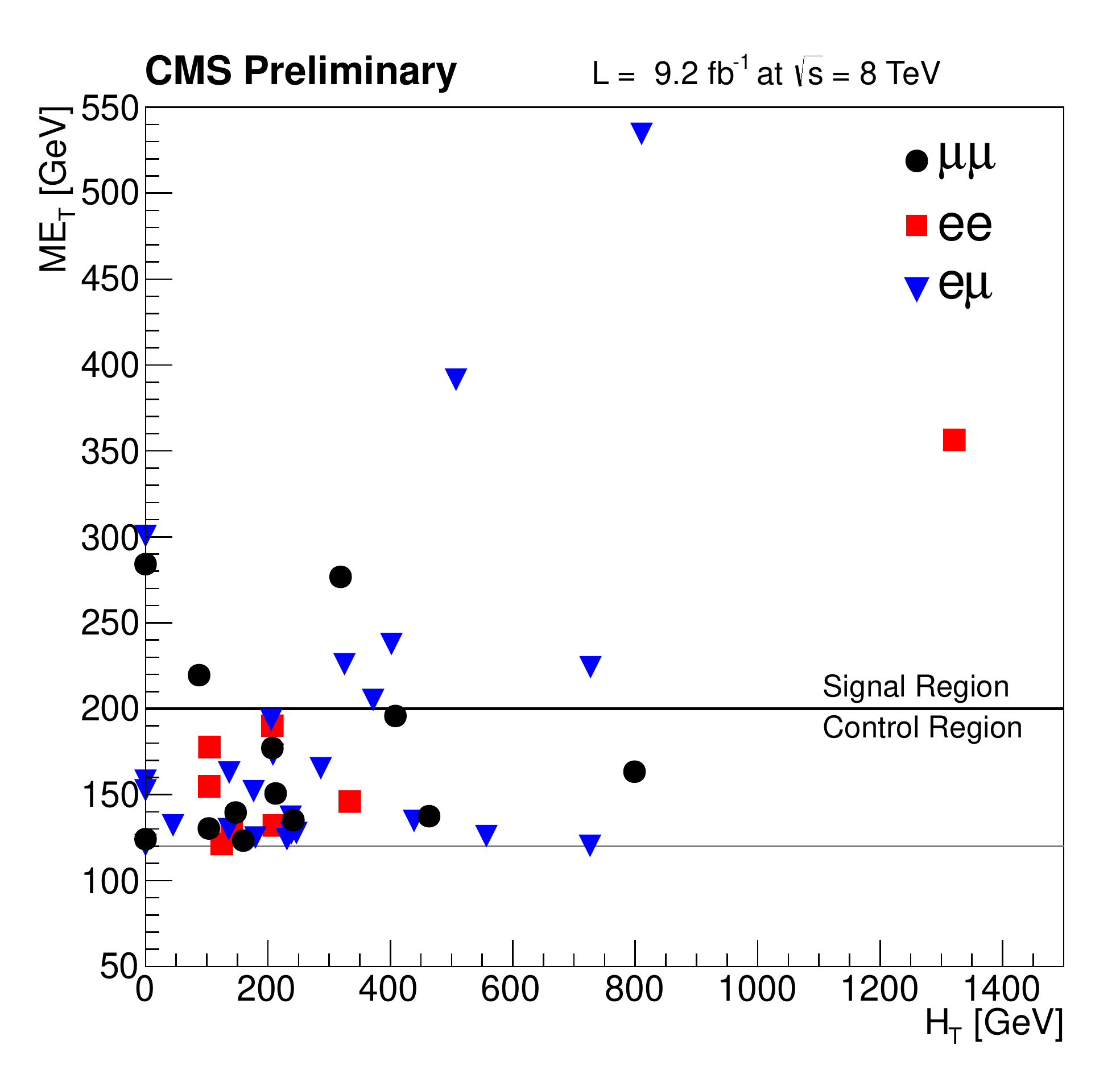}}
		\end{minipage}
	\end{center}
	\caption{Distribution of \MET versus \HT for the events passing the baseline selection. Left plot shows the events passing the selection for the analysis targetting strong production and the right plot the ones targeting electroweak production of SUSY.}
	\label{fig:baseline}
\end{figure}

Figure \ref{fig:baseline} shows all the event passing the selection in the \MET and \HT plane.

\section{Background estimation}
There are some sources of SM background to potential new physics signals: events with one or two fake leptons, opposite-sign events in which one of the electron charge has been badly measured and events with two same-sign prompt-leptons. 

A description of the relevance of these backgrounds and how they are estimated is presented in this section. The validity of these estimation methods is proved in the baseline regions, that are background dominated.

\subsection{Backgrounds with one or two fake leptons}
Backgrounds with one or two fake leptons, include processes such as semi-leptonic $\mathrm{t\bar{t}}$ or $W+jets$ where one of the leptons comes from a heavy-flavor decay, misindentified hadrons, muons from light-meson decay in flight, or electrons from unidentified photon conversions. We estimate this background starting from measuring the probability of a lepton being fake or prompt using a QCD or Z enriched sample respectively. We then apply those probabilities to events passing the full kinematic selection but in which one or two of the leptons fail the isolation requirements.
About 40-50\% of the total background is due to this processes and we assign a 50\% systematic uncertainty to account for the lack of estatistics in the control sample as well as the little knowledge we about about the control sample composition.


\subsection{Events with charge mis-identification}
These are events with opposite-sign isolated leptons where one of the leptons (typically an electron) and its charge is misreconstructed due to sever bremsstrahlung in the tracker materia (this effect is negligible for muons). We estimate this background by selecting opposite-sign $ee$ or $e\mu$ events passing the full kinematic selection, weighted by the probability of electron charge misassignment. This probability is measured in a $Z\rightarrow e e$ sample in data by simply calculating the ratio between same-sign and opposite-sign events in such sample and it validated in MC, this probability is of the order 0.02 (0.2)\% for electrons in the barrel(endcap).
This source of background only only accounts for the 5-10\% of the total background. A 20\% systematic uncertainty on this background is considered to account for the \pt dependence of the probability.  

\subsection{Rare SM processes.}
These include SM processes that yield two same-sign prompt leptons, including $\mathrm{t\bar{t}W}$, $\mathrm{t\bar{t}Z}$,  $\mathrm{W}^\pm\mathrm{W}^\pm$ among others. These processes constitutes about 30-40\% of the total background. $\mathrm{WZ}$ production is also very relevant for the EWK analysis, constituting nearly 40\% of the total background.  

All these background are obtained from Monte Carlo simualtions. For the $\mathrm{WZ}$ production the MC is validated in dataa and a 20\% systematic uncertainty to account for the differences. Other backgrounds are assigned a 50\% systematic uncertainty to this background sources as we have very little knowledge on the cross-sections.

\section{Results}
\subsection{Strong production of SUSY}
The search is based on comparing observed and predicted yields in 8 signal regions with different requirements motivated by various possible new physics models. The definition of these search regions, as well as the observed and predicted yields are shown in Table \ref{tab:RA5}. 

\begin{table}[h]
\caption{\label{tab:SRcomb} A summary of the combination of results for this search. For each signal region (SR), we show its most distinguishing kinematic requirements, the prediction for the three background  (BG) components as well as the total, and the observed number of events.}
\label{tab:RA5}
\vspace{0.4cm}
\begin{center}
\resizebox{0.9\textwidth}{!}{
\begin{tabular}{|l|c|c|c|c|c|c|c|c|c|}
	\hline
					&   SR0	& SR1 & SR2 & SR3 & SR4 & SR5 & SR6 & SR7 & SR8 \\ \hline
    No. of jets		& $\geq 2$	    & $\geq 2$		& $\geq 2$		& $\geq 4$		& $\geq 4$		& $\geq 4$		& $\geq 4$		& $\geq 3$		& $\geq 4$       \\
	No. of btags    & $\geq 2$		& $\geq 2$		& $\geq 2$		& $\geq 2$		& $\geq 2$		& $\geq 2$		& $\geq 2$		& $\geq 3$		& $\geq 2$       \\
	Lepton charges  & $++/--$  		& $++/--$		& $++$			& $++/--$		& $++/--$		& $++/--$		& $++/--$		& $++/--$		& $++/--$        \\
	$\MET$          & $> 0$ \GeV 	& $> 30$ \GeV	& $> 30$ \GeV	& $> 120$ \GeV	& $> 50$ \GeV	& $> 50$ \GeV	& $> 120$ \GeV	& $> 50$ \GeV 	& $> 0$ \GeV     \\
	$\HT$           & $> 80$ \GeV 	& $> 80$ \GeV	& $> 80$ \GeV	& $> 200$ \GeV	& $> 200$ \GeV	& $> 320$ \GeV	& $> 320$ \GeV	& $> 200$ \GeV	& $> 320$ \GeV   \\
\hline
	Charge-misID BG	& $3.35 \pm 0.67$ 	& $2.70 \pm 0.54$	& $1.35 \pm 0.27$  & $0.04 \pm 0.01$ & $0.21 \pm 0.05$ & $0.14 \pm 0.03$ & $0.04 \pm 0.01$ & $0.03 \pm 0.01$ & $0.21 \pm 0.05$	\\
	Fake BG 	  	& $24.77 \pm 12.62$ & $19.18 \pm 9.83$ 	& $9.59 \pm 5.02$  & $0.99 \pm 0.69$ & $4.51 \pm 2.85$ & $2.88 \pm 1.69$ & $0.67 \pm 0.48$ & $0.71 \pm 0.47$ & $4.39 \pm 2.64$  \\
	Rare SM BG      & $11.75 \pm 5.89$ 	& $10.46 \pm 5.25$ 	& $6.73 \pm 3.39$  & $1.18 \pm 0.67$ & $3.35 \pm 1.84$ & $2.66 \pm 1.47$ & $1.02 \pm 0.60$ & $0.44 \pm 0.39$ & $3.50 \pm 1.92$  \\
	\hline
	Total BG        & $39.87 \pm 13.94$ & $32.34 \pm 11.16$ & $17.67 \pm 6.06$ & $2.22 \pm 0.96$ & $8.07 \pm 3.39$ & $5.67 \pm 2.24$ & $1.73 \pm 0.77$ & $1.18 \pm 0.61$ & $8.11 \pm 3.26$  \\
	Event yield     & 43                & 38                & 14               & 1               & 10              & 7               & 1               & 1               & 9              	\\
\hline
\end{tabular}
}
\end{center}
\end{table}

None of the search regions shows any significant excess over the SM background predictions, therefore we set interpret the results in several physics models\cite{RA5}, for example Figure \ref{fig:interpretation} shows the exclusion regions for gluino-pair production decaying into on-shell stops. We are able to exclude gluino masses up to 1 \TeV with such models.  

\subsection{Electroweak production of SUSY}
This analysis is targeting \char\neut production decaying via sleptons. This process naturally gives three-lepton final states. However when the mass of the intermediate slepton is too close either to the \char or to the \LSP, the third lepton would be too soft and the event would be missed by the tri-lepton analysis. We can recover such events using the same-sign analysis. We will assume that the strongly interacting particles do not play a role in this scenario.

\begin{figure} 
	\begin{center}
		\begin{minipage}{0.45\linewidth}
			\centerline{\includegraphics[width=0.9\linewidth]{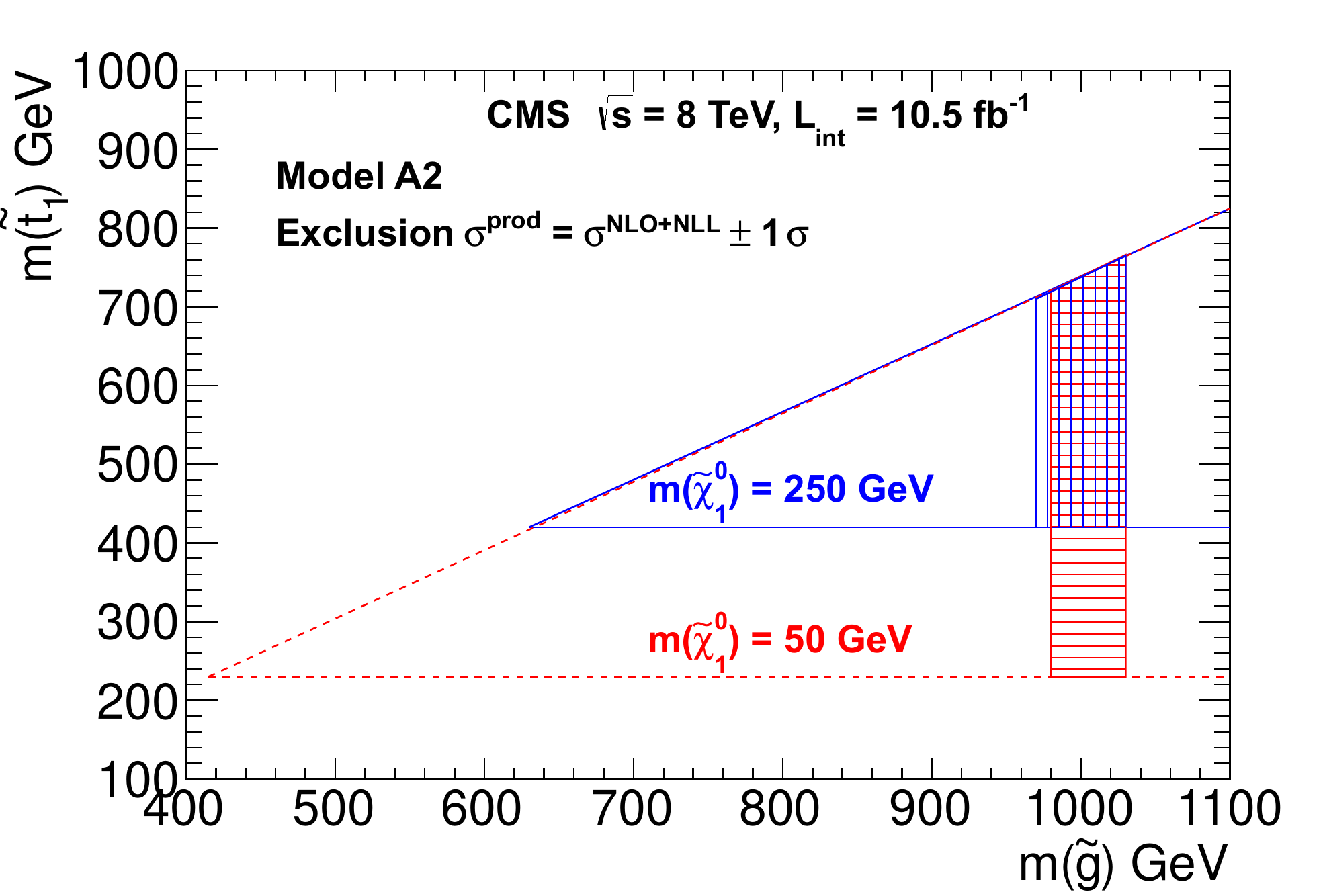}}
		\end{minipage}
		\begin{minipage}{0.45\linewidth}
			\centerline{\includegraphics[width=0.9\linewidth]{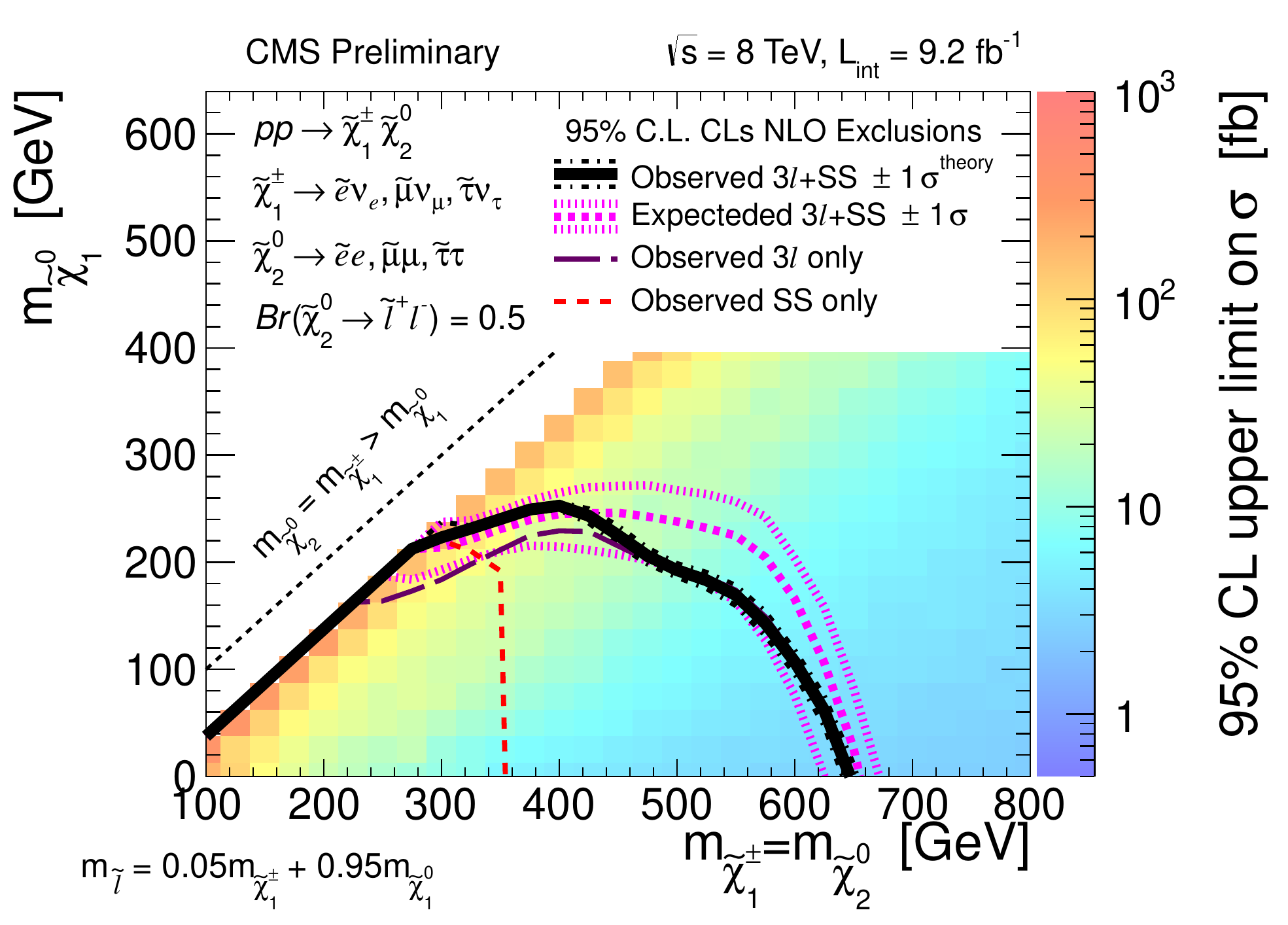}}
		\end{minipage}
	\end{center}
	\caption{Exclusion regions for gluino-pair production, in the $m(\widetilde{t}_1)$ vs $m(\widetilde{g})$ plane, where each of the gluinos decays $\widetilde{g}\rightarrow t\bar{t}\LSP$ with on-shell stops (left). Exclusion region for \char\neut production with intermediate sleptons assuming that the mass of the slepton is very close to the mass of the \LSP.}
	\label{fig:interpretation}
\end{figure}

Figure \ref{fig:interpretation} shows the exclusion region for \char\neut pair production decaying via sleptons, when the mass of the slepton is very close to the mass of the LSP. One can see that the same-sign analysis (red-dashed line) drives the exclusion near the diagonal. We  are able to exclude chargino/neutralino masses up to roughly 600 \GeV.

\section{Conclusions}
We have presented results of a search for new physics with events with same-sign dileptons using the CMS detector at the LHC. No significant deviations from the standard model expectations are observed. The results are used to set exclusion limits into several SUSY models, both assuming strong-dominated production and electroweak-dominated production. With the first we are able to probe gluino masses up to 1 \TeV and with the latter we exclude chargino/neutralino masses up to roughly 600 \GeV.

\section*{Acknowledgments}
We wish to congratulate our colleagues in the CERN accelerator departments for the excellent performance of the LHC machine. We thank the technical and administrative staff at CERN and other CMS institutes, and acknowledge support from: FMSR (Austria); FNRS and FWO (Belgium); CNPq, CAPES, FAPERJ, and FAPESP (Brazil); MES (Bulgaria); CERN; CAS, MoST, and NSFC (China); COLCIENCIAS (Colombia); MSES (Croatia); RPF (Cyprus); Academy of Sciences and NICPB (Estonia); Academy of Finland, MEC, and HIP (Finland); CEA and CNRS/IN2P3 (France); BMBF, DFG, and HGF (Germany); GSRT (Greece); OTKA and NKTH (Hungary); DAE and DST (India); IPM (Iran); SFI (Ireland); INFN (Italy); NRF and WCU (Korea); LAS (Lithuania); CINVESTAV, CONA- CYT, SEP, and UASLP-FAI (Mexico); MSI (New Zealand); PAEC (Pakistan); MSHE and NSC (Poland); FCT (Portugal); JINR (Armenia, Belarus, Georgia, Ukraine, Uzbekistan); MON, RosAtom, RAS and RFBR (Russia); MSTD (Serbia); MICINN and CPAN (Spain); Swiss Funding Agencies (Switzerland); NSC (Taipei); TUBITAK and TAEK (Turkey); STFC (United Kingdom); DOE and NSF (USA).


\section*{References}
\begin{thebibliography}{99}
\bibitem{CMS} CMS Collaboration, ``The CMS experiment at the CERN LHC'', JINST 3 S08004 (2008).
\bibitem{RA5} CMS Collaboration,``Search for new physics in events with same-sign dileptons and $b$ jets in $pp$ collisions at $\sqrt{s}=8$ TeV'', JHEP {\bf 1303} (2013) 037 [arXiv:1212.6194].
\bibitem{SUS12022} CMS Collaboration,``Search for direct EWK production of SUSY particles in multilepton modes with 8TeV data'',  CMS-PAS-SUS-12-022.

\end{thebibliography}
\end{document}